\documentclass{ieeeaccess}
\usepackage{cite}
\usepackage{amsmath,amssymb,amsfonts}
\usepackage{algorithmic}
\usepackage{graphicx}
\usepackage{textcomp}
\def\BibTeX{{\rm B\kern-.05em{\sc i\kern-.025em b}\kern-.08em
    T\kern-.1667em\lower.7ex\hbox{E}\kern-.125emX}}
\begin{document}
\history{}
\doi{}

\title{A hybrid prognosis approach \newline for robust lifetime control of commercial wind turbines}
\author{\uppercase{Edwin Kipchirchir}\authorrefmark{1}, 
\uppercase{Jonathan Liebeton \authorrefmark{2}, and Dirk S{\"o}ffker}.\authorrefmark{3},\IEEEmembership{Member, IEEE}}
\address[1]{Chair of Dynamics and Control, University of Duisburg-Essen, Lotharstr. 1-21, 47057, Duisburg, Germany (e-mail: edwin.kipchirchir@uni-due.de)}
\address[2]{Chair of Dynamics and Control, University of Duisburg-Essen, Lotharstr. 1-21, 47057, Duisburg, Germany (e-mail: jonathan.liebeton@uni-due.de)}
\address[3]{Chair of Dynamics and Control, University of Duisburg-Essen, Lotharstr. 1-21, 47057, Duisburg, Germany (e-mail: soeffker@uni-due.de)}
\tfootnote{This work is partly supported through a scholarship awarded to the first author by the German Academic Exchange Service (DAAD) in cooperation with the Ministry of Education of Kenya, for his Ph.D. study at the Chair of Dynamics and Control, UDE, Germany.}

\corresp{Corresponding author: Edwin Kipchirchir (e-mail: edwin.kipchirchir@uni-due.de).}

\begin{abstract}
Dynamic fluctuations in the wind field to which a wind turbine (WT) is exposed to are responsible for fatigue loads on its components. To reduce structural loads in WTs, advanced control schemes have been proposed. In recent years, prognosis-based lifetime control of WTs has become increasingly important. In this approach, the prognostic controller gains are adapted based on the state-of-health (SOH) of the WT component to achieve the desired lifetime. However, stochastic wind dynamics complicates estimation of the SOH of a WT. More recently, robust controllers have been combined with real-time damage evaluation models to meet prognosis objectives. Most rely on model-based online load cycle counting algorithms to determine fatigue damage, with analytical models providing the degradation estimate. However, most use load measurements that are either unreliable or unavailable in commercial WTs, limiting their practicality. In this contribution, a hybrid prognosis scheme combining data-driven load prediction and model-based damage estimation models for robust lifetime control of commercial WTs is proposed. A data-driven support vector machine (SVM) regression model is trained using loading data obtained from dynamic simulations using a $\mu$-synthesis robust disturbance accommodating controller (RDAC). The regression model uses available WT measurements to predic tower load. Based on this prediction, an online rain-flow counting (RFC) damage evaluation model estimates the damage level and lifetime of the tower. The RDAC controller gains are dynamically adapted to achieve a predefined damage limit and lifetime. The proposed approach is evaluated on a 5 MW reference WT and its performance is compared with a model-based prognosis scheme using ideal WT tower measurement. Results demonstrate the efficacy of the proposed approach to control the fatigue loading in WT components to achieve a predefined damage level and lifetime without compromising generator speed regulation.
\end{abstract}


\titlepgskip=-15pt

\maketitle

\section{Introduction}
\label{introduction}
Concerns related to global climate challenges have led to prioritization of renewable energy development \cite{IRENA2021}. Wind is the fastest growing renewable energy source, and is expected to remain so for the foreseable future \cite{GWEC2023}. In tandem with increasing demand for wind energy, commercial wind turbines (WTs) have increased in size and power rating. With larger, slender and heavier components, these utility-scale WTs are subject to increased structural loads, making them less tolerant to faults and performance degradation \cite{Gao2021}. Wind turbines are typically located in harsh environments, with unsteady variations of wind loads being responsible for their relatively high failure-rate \cite{Lei2019}. Fatigue loads in WT components are attributed to dynamic variation of the wind field. To ensure that commercial WTs operate within their design lifetime, advanced  multi-input multi-output (MIMO) control strategies have been developed to reduce structural loads, particularly in the large rotor blades and tower components. These control schemes include additional objectives such as power optimization and speed control. 

Wind turbines experience faults or failures due to component degradation through aging or extreme load events. This leads to system downtime and economic losses. To ensure operational safety and reliability of WT systems, structural health monitoring (SHM) and prognosis schemes for estimating and predicting degradation should be integrated to advanced MIMO control strategies. This ensures that the design lifetime WTs is achieved. Significant improvements in SHM has been achieved through online fault detection and condition monitoring (CM) \cite{Beganovic2018}. In CM, operational parameters of a system are continuously monitored to identify significant variations which indicate incipient faults. Therefore, CM of critical components seeks to detect significantly large deviations in SOH from a healthy state and taking measures to avoid catastrophic failure. Full or partial automatic diagnostic schemes detect and locate potential faults in a WT. This enables optimization of the maintenance procedure, reduces downtime and economic loss, and avoids complete failure \cite{Xiang2021}. Diagnosis and prognosis schemes are usually incorporated in SHM to enable decision-making on maintenance or replacement actions. These schemes are also used as a module in a prognostics control scheme. 

In recent years, prognosis-based lifetime control and extension of WTs has become increasingly important for ensuring power supply and operational reliability. In this context, a fatigue damage or degradation model is used to determine the consumed lifetime of a component. Subsequently, the prognostics controller gains are adapted based on the state-of-health (SOH) or remaining useful life (RUL) of the WT component to achieve the desired lifetime. During operation, almost all WT components are subjected to varying load combinations due to stochastic wind fields. Therefore, determination of the current SOH and prediction of the RUL of WT components is a challenge. Stochastic wind dynamics complicates estimation of the SOH in a WT component as it causes nonstationary load profiles. 

Various SHM and prognosis approaches for RUL prediction in WTs have been proposed. Physics/model-based prognosis methods use either physical or mathematical models of the degradation patterns to predict the RUL of critical components in real-time \cite{Rezamand2021}. Degradation models are obtained either through physical modeling of the degradation process or through experimental modeling, where system identification tools are used to obtain models from process measurements. Using monitoring data,these degradation models are used to predict the damage evolution and thus determine the RUL. The limitation of model-based approaches is that accurate physical modeling of the degradation process is difficult to achieve due to uncertain future operating conditions, measurement errors, and data variability. This leads to uncertainty in the predicted RUL \cite{Kim2016}. On the other hand, data-driven prognosis approaches rely on featured data of the degradation process rather than explicit input/output models, and suitable machine learning (ML) techniques to build a knowledge-base that represents an explicit dependency of system variables, and a degradation model enabling prediction of future SOH and RUL \cite{Ding2018,Gao2021,Badihi2022}. Data-driven approaches have been applied to prognosis of modern commercial WTs due to abundance of process monitoring data from high frequency (1 kHz) supervisory control and data acquisition (SCADA) systems \cite{Lin2020}. This historical data is used to learn the WT’s performance dynamics, estimate the SOH from real-time data, and predict its RUL. Data-driven prognosis approaches have a limited ability to learn complex signals with nonlinear characteristics, such as the nonstationary degradation patterns in WTs due to several failure modes \cite{Herp2018}. Although sufficient process data is required for accurate predictions, traditional data-driven methods have slow convergence speed and low prediction accuracy when processing large amounts of data \cite{Xiang2021}. To adress this problem, modern deep-learning methods have accelerated convergence speed and improved prediction accuracy. However, longterm dependencies hidden in sequential/timeseries data are not considered \cite{Lei2019}. 

In recent years, hybrid (model-based data-riven) prognosis approaches which capitalize of the unique advantages of each approach and compensate for their limitations have been proposed. They have good prediction performance due accurate modeling of uncertainty. However, hybrid prognosis algorithms can be very sophisticated and are constrained by the requirement for physical modeling of degradation events \cite{Bousebsi2023}. Therefore, hybrid models require a reliable physical degradation model and sufficient historical process data for successful implementation. Data-driven methods are not reliable for prognosis of WTs at the beginning of their operational life due to insufficient databases covering faulty scenarios as most of the collected data characterizes normal operatation. On the other hand, useful features covering both of these operations can be generated using physical models \cite{Djeziri2018}. High fidelity softwares for simulating several dynamics of commercial WTs can be used for developing these physics models. Because WTs are designed to operate for decades, hybrid prognosis approaches are suitable for RUL prediction of WTs over their entire lifetime.

Recent developments have seen the use of robust controllers combined with real-time damage evaluation models to achieve prognosis objectives in WTs. In \cite{Do2020} and \cite{Kipchirchir2023a}, adaptive robust lifetime control schemes for tower and blade fatigue life control, respectively, are proposed. These prognostics controllers rely on model-based online load cycle counting algorithms to determine fatigue damage accumulation with analytical models providing the degradation estimate. However, these rely on load measurements, which may be unavailable in commercials WTs, hence limiting their practicality. In addition, the methods are applied to a small 1.5 MW RWT which does not reflect the current trend in the size of commercial WTs.

In this contribution, a hybrid prognosis scheme that combines data-driven load prediction and model-based damage estimation models for robust lifetime control of WTs is outlined.  A support vector machine (SVM) regression model is selected for tower load prediction because it is a well-known ML method and can be directly integrated into the existing framework. The data-driven SVM regression model is trained and tested using timeseries featured data from dynamic simulations (for various wind conditions) using a collective pitch control (CPC)-based $\mu$-synthesis robust disturbance accommodating controller (RDAC), which is based on the previously proposed independent pitch control (IPC)-based RDAC controller \cite{Kipchirchir2024}. The regression model uses available WT measurements to predict the tower load. Based on this prediction, an online RFC damage evaluation model \cite{Musallam2012}, estimates the SOH and lifetime of the tower. Using the estimated lifetime and a set of predefined thresholds, the RDAC controller gains are dynamically adapted to achieve a predefined damage limit and lifetime. The proposed hybrid lifetime control approach is applied to the 5 MW national renewable energy laboratory (NREL) reference WT (RWT) \cite{Jonkman2009}. Its performance is compared with a model-based prognosis scheme that uses ideal WT tower measurement. The simulation results illustrate the efficacy of the proposed method in managing fatigue loading in WT components, ensuring attainment of a predetermined damage threshold and operational lifespan while maintaining generator speed regulation.

The paper is organized as follows. In Section \ref{sec:SHMP}, SHM and prognosis methods applied to WTs are discussed. In Section  \ref{sec:SHMP}  a review of available SHM and prognosis methods applied to WTs is given. In Section \ref{sec:lifetime}, application of SHM and prognosis in lifetime control of WTs is discussed. In Section \ref{sec:hybrid} the proposed hybrid prognosis approach for robust lifetime control is described. The $\mu$-synthesis RDAC controller, online load prediction, and damage evaluation models are also discussed. In Section \ref{sec:results}, results obtained from closed-loop dynamic simulation using the proposed approach applied to the NREL 5 MW reference WT are presented and discussed.  Lastly, summary and conclusions are given in Section \ref{sec:conclusion}.
\section{structural health monitoring and prognosis of wind turbine systems}
\label{sec:SHMP}
Wind turbines usually experience faults or failures due to component degradation through aging or ephemeral events \cite{Gao2021}. This leads to system downtime and economic losses for wind farm operators. Reduced reliability of components occurs due to change in their material properties with use. Abnormal behaviours of WTs are classified into faults or failures. While a fault is an unacceptable deviation measured parameters from the normal values, which could signal a damage to a system, a failure is a complete loss of functionality of the system \cite{Beganovic2016}. When a fault occurs, fault diagnosis is carried out to identify the type, location, criticality, and patterns of faults at an early stage. As part of SHM systems, appropriate maintenance actions are taken to correct detected faults. When an incipient fault in a WT is identified through condition monitoring (CM) or its RUL predicted by a SHM and prognosis system, the decisions to be made include shutdown, reducing loads by derating, or implementing maintenance \cite{Lei2018}. The maintenance actions taken may include corrective maintenance, which involves restoring the system to its undamaged state, emergency maintenance, which aims to avoid complete failure of components, or predictive maintenance, which relies on predicted future failure based on the current state-of-health (SOH), or reliability of the system to plan maintenance ahead of time \cite{Beganovic2016,Lei2018}. To achieve this, CM of system's health is required. 

Integration of CM systems (CMS) in WTs enables advanced techniques such as fault detection and diagnosis (FDD) and lifetime prognosis (LTP) \cite{Badihi2022}. They also enable cost-effective preventive maintenance, called condition-based maintenance (CBM). In CM, a systems operational parameters are continuously monitored to identify significant variations that indicate incipient faults. Therefore, CM of critical components seeks to detect significantly large deviations in SOH from a healthy state and take measures to avoid catastrophic failure. While CMS monitor key parameters including drive-train vibration, oil quality, component or system temperatures, etc., they are usually deployed as add-ons to the WT. However, a supervisory control and data acquisition (SCADA) system is included in every utility-scale WT for performance monitoring \cite{Pandit2023}. It is employed to collect parameters relevant to the operating characteristics of a WT including wind speed, power, current, voltage, temperature, etc. Due to its low-cost as no additional sensor installation is required, a wide variety of methods employing SCADA data for fault prognosis have been developed.

Fully- or semi-automatic diagnostic schemes detect and localize potential faults in a WT. This allows optimization of the maintenance procedure, reduces downtime and economic loss, and avoids complete failure \cite{Xiang2021}. To improve the reliability of a system, fault diagnosis is usually employed using either hardware or software/analystical redundancy \cite{Gao2015}. Hardware redundancies such as sensors and actuators can be implemented in WTs to automatically detect faults based on the residual signatures of collected data. However, this comes at a high price of increased data acquistion complexity, weight, and space requirements. On the other hand, software redundancies that use mathematical models to generate residuals overcome these limitations \cite{Habibi2019}. Because an analytical model of the system is used, software redundancy is also called model-based fault diagnosis \cite{Lei2019}. By extracting hidden features within SCADA data, the operating state of the WT can be identified and early failures can be detected and predicted \cite{Xiang2021}.

Prognosis is the process of predicting the progression of a monitored component from undamaged to failed state \cite{Gao2021}. Using the damage evolution or aging of a component, prognosis attempts to estimate the SOH and predict the RUL of the component before faults ultimately lead to failure. It predicts the progression of SOH from undamaged state of a WT component to when its structural reserves are consumed. Prognostic algorithms use long-term predictions to describe the evolution of a fault indicator, so as to estimate the time-of-failure (TOF), or RUL of a faulty component or system \cite{Rozas2020}. To achieve this, a thorough knowledge of the degradation processes, anticipated future loading profiles, and characteristics of all sources of uncertainty is required \cite{Jaramillo2022}. The RUL prediction made at any given time uses a mathematical model of the failure criterion adapted to online measurement data. Therefore, a robust technique should be used to predict an optimal RUL for the component under consideration based on a predefined failure criterion \cite{Rezamand2021}. Furthermore, uncertainties including unknown future operation conditions and damage process, should be considered in the RUL prediction with associated confidence interval. Based on varius fault diagnosis and CM strategies, WT health status can be assessed. Therefore, degradation patterns can be established, allowing for implementation a prognostic scheme for failure prediction \cite{Gao2021}. By measuring system performance degradation from normal operation, the RUL can be estimated through fault prognosis. Diagnosis and prognosis schemes are usually incorporated in SHM systems. Structural health monitoring and prognosis systems are employed as modules in reliability-oriented control to aid in decision-making about maintenance and replacement of components for lifetime control and extension \cite{Beganovic2016}. Fault prognosis techniques are broadly classified into physics/model-based, data-driven, and hybrid approaches \cite{Gao2021,Rezamand2021}. 

\subsection{Model-based prognosis approaches}
Model-based methods are suitable for fault prognosis of WTs in real-world operating conditions. They employ physical or mathematical models of the degradation trend, and leverage knowledge of the system's dynamical behaviour, CM data, and damage evaluation models to predict RUL of critical components in real-time \cite{Rezamand2021}. Model-based approaches are developed assuming that the failure process conforms to a physical law such as fatigue cracking, which obeys the Paris-Erdogan model \cite{Teng2020}. The required mathematical models, obtained using theoretical or experimental modeling approaches such as finite element methods and fatigue propagation models, respectively \cite{Ding2018}, are usually developed as degradation models to describe the degraded process for LTP \cite{Badihi2022}. Therefore, WT models obtained either through physical modeling of the degradation process based on mathematically formulated natural laws, or by experimental modeling, where the model is identified from the process measurements using system identification techniques that express the input-output relationship in a mathematical model. Measured monitoring data is used to identify model parameters, which are then used to predict future damage/degradation evolution, hence determining the RUL. The choice of which model to use is problem-specific. The degradation model is a function of loading conditions, elapsed time/cycle, and model parameters. Loading conditions and time are often assumed to be given \cite{Kim2016}.

Although model-based prognosis approaches require a relatively small amount of observed data to predict the future damage behaviour, the measured data must be directly related to the physical model. Furthermore, failure prediction of complex systems without well-defined physics describing the degradation process is challenging \cite{Kim2016}.  An accurate physical model that describes the damage degradation as a function of time does not exist in practice. In addition, the future usage condition is the most significant source of uncertainty. Imperfections in the degradation model are caused by uncertain future loading/operating conditions, measurement errors and noise from onboard sensors and actuators used to quantify damage growth, and variations associated with the data used in system identification. These cause uncertainty in the estimated model parameters and thus, the predicted RUL \cite{Kim2016}. Therefore, to improve the degradation model while taking future uncertainties into account, parameters of the model should be estimated in real-time using Bayesian algorithms such as the overall Bayesian method (BM) and recursive Bayesian (filtering-based) methods such as Kalman filter (KF), extended/uncented KF, and particle filter (PF) \cite{Badihi2022}. It relies on measurement data to reduce the uncertainty of model parameters. It is preferred over other parameter estimation techniques such as nonlinear least squares (NLS) and maximum likelihood estimation methods due to its ability to estimate the uncertainty of identified model parameters. Therefore, most model-based approaches are based on Bayesian inference \cite{Kim2016,Badihi2022}. Parameter estimation algorithms form the basis for classification of model-based approaches. For a degradation model that is a nonlinear function of model parameters, the regression process to find unknown parameters is known as NLS. The KF-family and PF are based on the filtering techniques that recursively update the parameters using one measurement data at a time. In contrast to PFs, which have no restrictions on systems and noise types, factors such as initial conditions and variance of the parameter as well as approximation in linearization affect the performance of KF-family \cite{Kim2016}. The choice on which filtering method to use is dictated by the dynamics of the system (linear/nonlinear) and the type of noise distribution (Gaussian or non-Gaussian) \cite{Saidi2017}. A summary of the model-based prognosis approaches reviewed in this work is given in Table \ref{Table1}.

\begin{table*}[h!]
\centering
\caption{Examples of existing literature on model-based prognosis of wind turbines}
\begin{tabular}{p{0.1\linewidth} p{0.2\linewidth} p{0.25\linewidth} p{0.25\linewidth}}
\hline
\textbf{Approach} & \textbf{BM} & \textbf{PF} & \textbf{BM+KF} \\
\hline
\textbf{References} &  \cite{Ding2018,Herp2018,Rezamand2020,Rezamand2021} & \cite{Fan2015,Saidi2017,Valeti2019} & \cite{Teng2020,Wang2020,Jaramillo2022}  \\
                  
\hline
\textbf{Monitored components} & Gearbox, bearing  & Drivetrain, blade  & Bearing, rotor blades  \\
                                            
\hline
\textbf{Advantages} & \multicolumn{3} {c} {Have excellent real-time capability due to their online implementation}  \\
                   & \multicolumn{3} {c}{Require a relatively small amount of observed data to predict future damage} \\
\hline
\textbf{Limitations} & \multicolumn{3} {c} {Failure prediction of complex systems without well-defined physics describing the degradation process is challenging}   \\
& \multicolumn{3} {c} {Measured data must be directly related to the physical model}   \\
                 
\hline
\end{tabular}
\label{Table1}
\end{table*}

Model parameters of the overall Bayesian approach are estimated as a posterior probability distribution function, which is proportional to the product of the likelihood of the observed data and the prior probability distribution \cite{Choi2010}. The main challenge of this method is choosing the right options for the sampling process. Markov chain Monte Carlo (MCMC) method is the most effective sampling method \cite{Kim2016}. Bayesian inference-based methods are proposed in \cite{Ding2018,Herp2018,Rezamand2021} for fatigue life prediction of WT drive-train components. In \cite{Ding2018}, an approach which integrates gear physics models and health condition data for RUL prediction of a WT gearbox under instantaneously varying wind load conditions is developed. The focus of the work is on prognosis of fatigue crack at the root of a spur gear. The uncertain distribution of material parameter in the modified Paris crack propagation model is estimated using a Bayesian inference based on CM data at each inspection interval and the predicted failure time is updated. Improved prediction accuracy is claimed compared to a constant-load method. Prognosis methods for WT bearings are proposed in \cite{Herp2018,Rezamand2020,Rezamand2021}. Wind turbine bearing failure prediction is proposed in \cite{Herp2018} by abstracting bearing residual states using a Bayesian inference and Gaussian processes. The model is trained using run-to-failure bearing timeseries, assuming that the feature space is described by a multi-variate Gaussian process. A high failure prediction horizon of approximately a month is achieved for bearing over-temperature. In \cite{Rezamand2020}, an adaptive Bayesian algorithm for RUL prediction of faulty WT bearings based on extracted features is developed. To improve RUL prediction accuracy a fusion strategy based on an ordered weighted average operator is used. In \cite{Rezamand2021}, estimation of RUL of WT drivetrain bearings via an adaptive Bayesian algorithm based on failure dynamics and prevailing operating conditions is proposed. The influence of environmental conditions such as ambient temperature and wind speed on bearing failure dynamics is defined by SCADA data and vibration signals. The method provides a higher RUL prediction than a naive Bayesian algorithm. 

The Bayesian framework is a probabilistic algorithm for sequential state estimation using Bayes rule, which is suitable for solving the problem of characterizing the posterior probability density function (PDF) of the state vector in dynamical systems \cite{Jaramillo2022}. For a dynamical system which is linear with Gaussian disturbances or smooth nonlinearity, the Bayesian framework problem can be solved using the KF, extended KF or unscented KF. However, in practice state dynamics are usually nonlinear, time-varying, and affected by non-Gausssian disturbances. This requires PFs to compute the suboptimal solution of the Bayesian framework. Also known as sequential Monte Carlo method, a PF is the most popular model-based technique for prognosis \cite{Kim2016}. It has a similar framework to recursive Bayesian update. A PF aims to represent the posterior PDF of the state vector at each time step using a set of random samples and weights called particles/samples \cite{Jaramillo2022}. In fault diagnosis and prognosis, PFs are used for online estimation of states which are unmeasurable or would require costly sensors. State estimation can also be performed for nonlinear processes with non-Gaussian uncertainties \cite{Patwardhan2012}. If one of the states in the state vector is a SOH indicator, the magnitude of incipient faults can be estimated. Fault isolation can also be realized if different fault modes present in the state vector are estimated \cite{Jaramillo2022}. Uncertainties associated with PF-based fault prognosis may arise from the model, future usage/load profile, and the prognostic algorithm. Approaches for WT drivetrain prognosis using PF are proposed in \cite{Fan2015,Saidi2017}. In \cite{Fan2015}, RUL prediction of a WT gearbox is implemented. To validate the approach, measurement data from a SCADA system for a WT gearbox is used. Moreover, the model is verified using real-time data, with results showing the practical value of the approach. A PF-based approach for prognosis of track degradation on a WT high speed shaft bearing (HSS) is proposed in \cite{Saidi2017}. Validation results on real drivetrain data show that the method outperforms both SVR and Kalman smoother approaches. In \cite{Valeti2019}, PFs are used to estimate the RUL of a fatigue damaged WT blade subjected to turbulence. The PDFs for RUL of WT blades are estimated for diffident initial crack sizes, while PF is used for predicting fatigue damage evolution by consdiering nonlinearity and uncertainty in crack propagation.

Approaches based on integrated Bayesian framework and PF are proposed in \cite{Teng2020,Wang2020,Jaramillo2022}. To solve the nonlinear tracking problem, uncented PF (UPF), which combines unscented Kalman transform and PF is used \cite{VanDerMerwe2001}. An approach based on improved UPF is developed in \cite{Teng2020} for RUL prognosis of WT bearings. Its practicality for field use is attributed to its high dependence on measurement data rather than on parameters of the initital degradation model. The efficacy of the approach is verified using three life-cycle bearings from an on-site WT. A combined diagnosis and prognosis approach for predictive maintenance of WT bearings with limited degradation data is developed in \cite{Wang2020}. Bearing incipient fault signatures are diagnosed using wavelet transform, and an algorithm is used to represent the bearing defects. This feature and the physics behind this degradation process is modeled in a Bayesian framework. A PF is used for online estimation of model parameter and prediction of RUL with recursive quantification of the uncertainty. Validated results using real WT bearing aging data as in \cite{Herp2018} demonstrated the significance of the approach compared with a data-driven technique. In \cite{Jaramillo2022}, a robust monitoring for online damage diagnosis and prognosis of WT blades is proposed. Identified modal frequencies of the rotor blades are selected as a SOH indicator to quantify the damage level, while features extracted from vibration signals are used to obtain inputs for a Bayesian framework based on PF. The algorithm generates long-term predictions of the SOH to estimate the TOF probablity mass function for the monitored blade. However, in contrast to \cite{Herp2018,Wang2020}, experimental data from fatigue tests is used for validation.

Spatio-temporal is wind speed, subjects WTs to varying loads, which lead to fatigue loading of structural components during operation. A number of models have been proposed for lifetime prediction, including fatigue life and progressive damage models, probabilistic damage growth models, and those based on virtual fatigue estimators. To estimate the lifetime from given fatigue stress data, rain-flow counting (RFC) algorithm is used in combination with Palmgren-Miner rule of linear damage accumulation and material-specific fatigue stress amplitude (S) vs. cycles number (N) curve, well known as S-N curve \cite{Cetrini2019}. The S-N curve defines the allowable fatigue cycles to failure. Fatigue life models calculate the RUL of a WT through extrapolation of fatigue data. Variables describing component deterioration are used in progressive damage models to estimate damage. In \cite{Mehlan2022,Moghadam2022}, RUL prognosis of offshore WT drivetrains based on digital twin concept is realized. A virtual sensor using a digital twin framework is proposed in \cite{Mehlan2022} for RUL assessment of WT gearbox bearings. The virtual sensor combines data from CMS and SCADA systems, with a physics-based gearbox model. Load estimation is realized using KF, least squares, and quasi-static methods. Accumulated damage obtained from a Palmgren-Miner model is employed as a SOH indicator for RUL assessment. In \cite{Moghadam2022}, a digital twin consisting of a torsional dynamic model, online measurements, and fatigue damage model, is used for RUL estimation. The approach is successfully evaluated on a 10 MW WT.

Although nonlinear models accurately represent the WT behaviour, most model-based fault diagnosis and prognosis methods have been developed for linear models despite the inconsistency in behaviour that exists between the linearized model and the highly nonlinear WT \cite{Habibi2019}. To increase model accuracy, nonlinear WT modeling frameworks such as linear parameter varying and fuzzy Takagi-Sugeno prototypes, both of which use a set of linear models, have been developed \cite{Habibi2019}. Several high-fidelity reference WT toolchains such as OpenFAST developed by NREL  \cite{OpenFAST2021}, horizontal axis wind turbine simulation code 2nd generation (HAWC2) developed by the Aeroelastic Design Research Program at Technical University Denmark (DTU) \cite{Larsen2007}, among others, are used by researchers and WT manufacturers. These have facilitated the development of SHM and prognosis schemes for WTs. 

\subsection{Data-driven prognosis approaches}
Also known as knowledge-based fault diagnosis and prognosis, this approach relies on featured data containing system degradation process instead of explicit input-ouput models, and integrates them to a suitable ML technique or statistical model to establish a knowledge-base which represents an explicit dependency of system variables, hence enables prediction of future conditions \cite{Ding2018,Lei2019,Jihin2019,Gao2021}. Therefore, the degradation model is based on black-box modeling methods such as NN \cite{Badihi2022}. With the aid of a classifier, a diagnostic decision can be arrived at by comparing the operational behaviour of the real WT system with the knowledge-base. The advantage of using data-driven approach for lifetime modeling of WTs lies in its ability to work with insufficient information from process data as well as its scalability and rapid deployment for various industries \cite{Jihin2019}. Data-driven approaches are widely used in practice because physical degradation models are rare \cite{Kim2016}. Wind turbines are instrumented with SCADA systems, whose timeseries signals such as vibration and acoustic measurements are used in fault diagnosis and prognosis.

In the absence of a physical model describing the degradation process, model-based prognosis may not be suitable. However, data-driven prognosis relies on observed data, not necessarily related to the degradation process, to identify patterns and predict future state \cite{Kim2016}. Although a physical model is not used, data-driven methods use mathematical models that are unique to the monitored system. While degradation data is required by model-based approaches for parameter estimation, aging data as well as observed data is used to train mathematical models in data-driven methods \cite{Kim2016}. Data-driven fault diagnosis and prognosis approaches have a limited ability to learn complex signals with nonlinear characteristics. When used in processing big data, traditional data-driven methods exhibit slow convergence speed and low prediction accuracy \cite{Xiang2021}. Athough modern deep-learning methods have accelerated convergence speed and improved prediction accuracy, most do not consider long-term dependencies hidden in sequential/timeseries data \cite{Lei2019}. 

Historical WT performance data is used in ML to learn the performance dynamics of the WT, estimate the SOH from real-time data, and predict its RUL. Timeseries measurement data is useed for fault diagnosis and prognosis of WTs as long-term dependencies hidden in this data is essential for generating classifiable features \cite{Lei2019}. Based on the data extraction process, data-driven fault diagnosis and prognosis can be classified into qualitative and quantitative approaches. Examples of qualitative approaches include, root-cause and fault tree analysis. Quantitative data-driven methods widely used in prognosis can be classified into statistical or nonstatistical (probabilistic or artificial intelligence) approaches \cite{Kim2016,Lei2019}.  A summary of the data-driven prognosis methods reviewed in this work is given in Table \ref{Table2}.

\begin{table*}[h!]
\centering
\caption{Examples of existing literature on data-driven prognosis of wind turbines}
\begin{tabular}{p{0.1\linewidth} p{0.15\linewidth} p{0.15\linewidth} p{0.15\linewidth} p{0.15\linewidth}}
\hline
\textbf{Approach} & \textbf{SVM} & \textbf{ANN} & \textbf{LSTM} & \textbf{PSO-ANFIS} \\
\hline
\textbf{References} &  \cite{Xiao2019,Li2020} & \cite{Eroglu2019,Ogliari2021} & \cite{Desai2020,Xiang2021,Vidal2023} &  \cite{Adedeji2021,Gougam2021} \\
                  
\hline
\textbf{Application} & HSS bearing, blade pitch system, power forecasting  & Power forecasting, fault prediction   & Main bearing, gearbox and generator bearings & Power forecasting, bearing\\
                                            
\hline
\textbf{Advantages} & \multicolumn{4} {c} {Rely on measured data not necessarily related to the degradation data}  \\
                   & \multicolumn{4} {c}{Ability to work with insufficient information from process data} \\
\hline
\textbf{Limitations} & \multicolumn{4} {c} {Require abundant data with many degradation sequences for training }   \\
& \multicolumn{4} {c} {Traditional approaches require long training and validation time} \\
& \multicolumn{4} {c} {Most do not consider longterm dependencies hidden in sequential/timeseries data}   \\
                 
\hline
\end{tabular}
\label{Table2}
\end{table*}

Common nonstatistical analysis approaches used for CM and fault diagnosis and prognosis in WTs include neural network (NN) and fuzzy logic (FL). In FL, which is inspired by human reasoning, a feature space is partitioned into fuzzy sets and  fuzzy rules are then applied for reasoning. Some of the statistical data-driven methods include principal component analysis (PCA), independent component analysis (ICA), Fisher discriminant analysis (FDA), subspace aided approach (SAP), and support vector machine (SVM). Using dimensionality reduction to preserve crucial trends in the original dataset for successful fault extraction, PCA, ICA, SAP, and FDA are used. On the other hand, SVM, a nonparametric method is used to detect faults in WTs due to its supreme classification capability. 

To improve reliability and accuracy of identification, statistical analyis-based methods can be coupled with suitable nonlinear kernels. In \cite{Saidi2017}, vibration-based prognostic scheme using SVM regression (SVR) combined with spectral kurtosis-derived time domain indices is proposed for prognosis of WT HSS bearings. A novel area under spectral kurtosis is used as a health indicator to determine rolling bearing fault and an SVR model is trained for lifetime prognosis. The proposed method is promising for early failure detection and estimation of degradation trend. Approaches based on SVM are proposed in \cite{Xiao2019,Li2020}. In \cite{Xiao2019}, diagnosis and prediction of faults in a WT pitch system is realized using radar chart and SVM. Indicator data obtained from a SCADA system is used in constructing radar charts which correspond to normal and faulty operations of WT. Features from the radar charts are used for SVM prediction. The proposed method returns a higher accuracy than an SVR model. An SVM is developed in \cite{Li2020} for short-term forecast of WT power production. To select the optimal parameters for SVM, an improved dragonfly algorithm with an adaptive learning factor and a differential evolution strategy is used. The method shows superior prediction performance compared to a back-propagation NN and a Gaussian process regression (GPR).

Intelligent diagnosis involves feature extraction and fault classification. To classify faults, traditional methods such as SVM rely on suitable preselected features. On the other hand, modern deep learning algorithms such as multi-layer perceptron (MLP), convolutional NN (CNN), and recurrent NN (RNN) rely on hierarchical architectures with multiple nonlinear layers to obtain generalizable features from large scale training data \cite{Lei2019}. Neural networks are the most common algorithms used for prognosis \cite{Kim2016}. Artificial neural networks (ANNs) are proposed in \cite{Eroglu2019,Ogliari2021}. In \cite{Eroglu2019}, an ANN is trained for fault prediction in WTs using a novel training algorithm called Antrain ANN. In \cite{Ogliari2021}, wind power prediction with a 24-hour horizon is proposed using ANN and physics models, and a hybrid of both. The dataset used for training included two-year hourly measurements from a wind farm. A comparison of the prediction accuracy of the models used showed the superiority of ANN over the physics model, while the hybrid model had the best overall performance.

To capture long-term dependencies hidden in timeseries data, long short-term memory (LSTM) model based on RNN uses recurrent behaviour and gates systems to learn features directly from multivariate timeseries signals. This can be combined with other ML methods to discover longer patterns leading to improved fault diagnosis and prognosis. In addition to a hidden state vector used in RNNs, LSTM includes a memory cell consisting of three gates, including the input, output and forget gates. It encodes the memory of observed information. In \cite{Desai2020,Xiang2021,Vidal2023}, approaches using LSTM and SCADA data for WT prognosis are proposed. In \cite{Desai2020}, an LSTM network is used to predict WT gearbox bearing failures caused by axial cracking based on one month of timeseries data. However, the optimal time window for onset of failures could not be determined. In \cite{Xiang2021}, a method combining CNN and LSTM is proposed for fault detection and prediction of WT gearbox and generator bearings using SCADA data analysis. The CNN cascades to LSTM based on an attention mechanism (AM). The AM is used to enhance important information by assigning different weights to LSTM to improve its learning accuracy through mapping weight and parameter learning. Predictive maintenance of WT main bearings using LSTM and SCADA variables of rotor speed, generated power, and temperature is proposed in \cite{Vidal2023}. Target failure can be detected up to four months in advance, giving operators sufficient time to make informed maintenance decisions.

When measured data is stochastic in nature, it is challenging to establish the link between RUL and the degradation indicator due to uncertainties from operating conditions and existence of multiple fault mechanisms. In this case, probabilistic methods are used to make predictions about future failures. Probabilistic approaches that have been used for multi-state degradation include Markov Model, Weiner process, Bayesian network, and proportional hazard model \cite{Cerrada2018}. A Bayesian network consists of nodes that correspond to random variables. The nodes are interconnected using conditional dependencies and can take distinct states. On the other hand, Markov models are used to estimate probabilities of future failures. This is achieved by finding the probabilities of each state as well as those associated with state transitions. Future states only depend on prior states \cite{Rezamand2020a}. Given that WT components exhibit multi-state degradation due to varied operating conditions, aging, and other factors, probabilistic prognosis methods can be used for accurate prediction of RUL. 

To overcome the limitations of each ML method and leverage individual advantages, data-driven prognosis methods can be merged to complement each other. An example of this is the adaptive neuro-fuzzy inference system (ANFIS), which takes advantage of the NN's extensive expert knowledge of system behaviour available in datasets and required by FL systems. Due to the black-box data processing structure in NN, back-tracking of output is difficult, resulting in slow convergence. Due to their accuracy, reduce computational time and robustness in searching for global optimal values of model parameters, hybrid ANFIS approaches optimized using particle swarm optimization (PSO) are proposed in \cite{Adedeji2021,Gougam2021}. In \cite{Adedeji2021}, very short-term power output forecasting is realized using two hybrid models of ANFIS, including PSO and generatic algorithm (GA), with PSO-ANFIS returning a higher forecast accuracy. In \cite{Gougam2021}, a PSO-ANFIS approach is proposed for modeling nonlinear degradation of extracted features to predict the RUL of a WT bearing using vibration signal. 

\subsection{Hybrid prognosis approaches}
\label{subsect:Review_SHM_hybrid}
By combining one or more of the aforementioned approaches for fault diagnosis and prognosis, hybrid techniques capitalize on the advantages of different methods while counteracting their individual limitations. Although model-based approaches require complex predefined physics or analytical models, they have excellent real-time capability due to their online implementation. In addition, model-based methods make use of fault information from measured data and empirical knowledge for reliable fault prognosis \cite{Pan2020}. From an economics perspective relative to wind farms aiming at cost reduction, it is not practical to install sensors in every component that needs to be monitored. Therefore relying on available SCADA measurements instead of CMS for fault prognosis is a cost-effective solution \cite{Pandit2023}. Data-driven methods rely on past operational data, usually acquired using SCADA and ML algorithms to build a knowledge-base used for CM and prediction of future degradation patterns. However, they require abundant historical lifecycle data of many degradation sequencies from similar sensors/actuators to train the model \cite{Wang2020,Rezamand2020}. In practice, most WTs have only limited degradation data available, which is not only costly and but also time-consuming to acquire. The degradation patterns are also nonstationary due to different failure modes and operating conditions \cite{Herp2018}. Data quality also affects the prediction accuracy of data-driven models \cite{Pan2020}. Furthermore, the training and validation process required for data-driven algorithms is time-consuming \cite{Gao2021}. 

In recent years, hybrid approaches for fault diagnosis and prognosis in WTs have been proposed. These techniques have good prediction performance because they accurately model of uncertainty. However, hybrid prognosis algorithms can be extremely sophisticated and are constrained by the requirement for physical modeling of degradation events \cite{Bousebsi2023}. Therefore, for successful implementation of a hybrid model, the physical degradation model should be reliable and sufficient historical degradation data should be available. A summary of the hybrid prognosis approaches reviewed in this work is given in Table \ref{Table3}.

\begin{table*}[h!]
\centering
\caption{Examples of existing literature on hybrid prognosis of wind turbines}
\begin{tabular}{p{0.1\linewidth} p{0.25\linewidth} p{0.25\linewidth} p{0.25\linewidth}}
\hline
\textbf{Approach} & \textbf{Physics-informed RNN} & \textbf{DBN+PF} & \textbf{ANFIS-PF} \\
\hline
\textbf{References} &  \cite{Yucesan2020} & \cite{Pan2020} & \cite{Cheng2018,Cheng2019}  \\
                  
\hline
\textbf{Application} & Main bearing  & Gearbox   & Gearbox, generator \\
                                            
\hline
\textbf{Advantages} & \multicolumn{3} {c} {Capitalize on the advantages of different methods while counteracting their individual limitations}  \\  
& \multicolumn{3} {c} {Good prediction performance because they enable accurate modeling of uncertainty}  \\                      
\hline
\textbf{Limitation} & \multicolumn{3} {c} {Algorithms can be extremely sophisticated and are constrained by the requirement for physical modeling of degradation events}   \\

\hline
\end{tabular}
\label{Table3}
\end{table*}

Hybrid physics-informed NN models are proposed in \cite{Yucesan2020,Pan2020} for prognosis in WTs. In \cite{Yucesan2020}, the fatigue life of WT main bearing, which is typically influenced by lubricant condition, is modeled by incorporating RNN into a lubricant degradation model. The approach gives accurate fatigue life prediction of WT main bearings. In \cite{Pan2020}, RUL prediction of WT gearbox is realized using a hybrid approach based on deep belief network (DBN) and improved PF. The DBN is used to denoise and merge vibrations to obtain the SOH indicator, while the PF is used for RUL prediction. A Wiener model is used to characterize the randomness of WT gearbox degradation operation, hence improving RUL prediction efficiency. To validate the effictiveness of the approach, simulated and experimental vibration signals from a WT gearbox are used. 

Bearing failure is the main cause of WT gearbox failures. Therefore, accurate prediction of RUL for gearboxes is critical for preventive maintenance. A WT drivetrain gearbox is a complex multi-component system, usually operating under varying load conditions. This makes it hard to obtain an accurate physical degradation model, especially for gearbox bearings. To solve this problem, approaches based on ANFIS-PF are developed in \cite{Cheng2018,Cheng2019}. In \cite{Cheng2018}, analysis of one phase stator current of a generator connected to a gearbox is used for prognosis and RUL prediction of WT gearboxes. The approach is realized using ANFIS for learning the fault feature state transition, and a PF algorithm is used to continuously predict the RUL based on the learned state transition and the new fault feature. To enhance the prediction performance of PFs by eliminating particle impoverishment in the resampling procedure due to low particle density, a particle modification method and improved multinomial resampling is proposed in \cite{Cheng2019}. On the other hand, an ANFIS is used to learn the state transition function in the fault degradation model using the SOH indicator obtained from monitoring data. The approach was evaluated on a doubly-fed induction generator (DFIG) of a 2.5 MW WT. 

Wind turbines are designed to operate for decades. However, there is insufficient data at the beginning of their operational life. Therefore, databases do not cover the useful features needed for successful fault prognosis using data-driven methods because collected data charaterizes only normal system operation. To solve this, physical models can be used to generate useful features covering both normal and faulty operation \cite{Djeziri2018}. Therefore, hybrid models are suitable for fault prognosis of a WT over its entire life.

\section{Application of structural health monitoring and prognosis in lifetime control of wind turbines}
\label{sec:lifetime}

Commercial WTs are less tolerant to performance degradation and unplanned downtime. The SOTA in lifetime control and extension strategies for these systems is the use of resilient or FTC \cite{Acho2016,Azizi2019,Elmaati2020,Jain2020} to minimize the impact of unanticipated faults or unexpected dynamics by maintaining the operation of a WT under a tolerated performance degradation. However, FTC control is reactive as it relies only on detected faults and does not address the problem of controlling life consumption in WT components to avoid fatigue failures while ensuring other control objectives such as power maximization are achieved. Therefore, to compensate for faulty components, the WT is operated with restricted power output until repairs are made and normal operation is restored. This is undesirable considering a 20-year lifespan of a WT.

From the foregoing discussion, it has been shown that SHM and prognosis approaches are useful in establishing the SOH and predicting RUL of WTs. In recent years, integration of SHM and prognosis in control of WTs has attracted attention in the research community. Performance and reliability of any given system is affected by its SOH. Therefore, prognosis of SOH or RUL of WTs is useful in developing health-oriented control strategies for optimal performance. This concept of continuously optimizing the control strategy based on the SOH was first introduced by \cite{Soeffker1997}. In the realm of WTs, the main trade-off is related to lifetime extension and power maximization.  

Within a wind farm, WTs impact each on power generation and structural loads through their wakes. Therefore, control strategies for mitigating wake effects are required. In \cite{Kanev2018,Vali2019} approaches for lifetime control of wind farms are proposed. The benefits of active wake control for lifetime power production and fatigue loading in real commercial wind farms is extensively studied in \cite{Kanev2018}. Active power control is employed in \cite{Vali2019}, to extend the service life of highly loaded WTs in a waked wind farm. The power reference signal of the entire wind farm is taken into account and fatigue loads in WTs are alleviated. Optimization of the control problem is based on a data-driven fatigue load model such that the lifetime tower fore-aft (F-A) loads of the WTs operating within the wind farm are balanced. In \cite{Cetrini2019}, an approach based on a SMC controller for online fatigue life alleviation in WTs is developed. An online fatigue estimator is employed as a virtual sensor of fatigue damage, which is fed to the controller to reduce fatigue of WT components. The approach is validated on the 5 MW NREL RWT.

In \cite{Pettas2018,Njiri2019}, approaches for lifetime extension of WTs using IPC-based MIMO controllers are proposed. In \cite{Pettas2018}, the trade-off between blade flap-wise (F-W) load mitigation and pitch actuation for extending the lifetime of a 10 MW RWT is investigated.  In \cite{Njiri2019}, a WT lifetime extension scheme incorporating an online RFC damage evaluation model and a IPC-based MIMO controller is developed. Based on the accumulated damage the controller is continuously adapted to trade-off between power production and tower F-A load mitigation to extend the service life. In \cite{Do2020}, an adaptive lifetime controller is proposed to achieve the desired lifetime of the tower. Depending on the damage accumulation and the predicted lifetime provided by an online damage evaluation model, the weights of the lifetime controller are varied. A health-oriented strategy for power control of a WT drive-drive to maximize its economic return over its entire lifecycle is proposed in \cite{Chen2022}. A model-based approach is employed for RUL prediction of the WT's power converter. A receding horizon model predictive control is employed for extending the converter fatigue life.

Although hybrid SHM and prognosis approaches are widely used in SOH and RUL prediction of WTs, little has been reported on the use of these approaches for lifetime control of WTs. Modern utility-scale WTs are instrumented with high frequency (1 Hz) SCADA systems \cite{Lin2020}. Additionally, model-based high fidelity toolchains have been developed to simulate the different dynamics of commercial WTs. Therefore, by taking advantage of abundant data generated by modern SCADA systems and combining this with the model-based softwares based on the digital twin concept, hybrid prognosis approaches can be developed for improved lifetime control and extension of WTs.

\section{Hybrid prognosis for robust lifetime control}
\label{sec:hybrid}
In this section, a hybrid approach for robust lifetime control of WTs is presented. An SVM regression model used for tower load prediction is combined with an online RFC damage accumulation model \cite{Musallam2012}. The SVM regression model is trained and tested using timeseries data obtained from closed-loop simulation using a $\mu$-synthesus RDAC controller. The effectiveness of the proposed approach is evaluated online on the 5 MW NREL RWT \cite{Jonkman2009}.

\subsection{Wind turbine model }
The land-based 5 MW NREL RWT \cite{Jonkman2009}, which is available in the high fidelity open-source fatigue, aerodynamics, structures, and turbulence (OpenFAST) software \cite{OpenFAST2021}, is used for the design and evaluation of the closed-loop coupled dynamic response of the proposed hybrid prognosis scheme. In Table \ref{Table4} a summary of the specifications of the 3-bladed WT is given. The 5 MW RWT model has 16 degrees of freedom (DOFs) describing the blades, tower, drive-train, generator, and nacelle motions. However, a few DOFs are enabled to capture the most important dynamics corresponding to the desired closed-loop performance in terms of structural load mitigation and generator speed regulation.
 
 \begin{table}[h!]
 \centering
 \caption{5 MW NREL reference WT specifications}
\begin{tabular}{lcc}
\hline
\textbf{Parameter}	& \textbf{Value} & \textbf{Unit}\\
\hline
Rated power  &   5  & MW\\
Hub height & 90 & m \\
Cut-in, Rated, Cut-out wind speed &  3, 11.4, 25  & m/s \\
Cut-in, Rated rotor speed	& 6.9, 12.1 & rpm\\
Gearbox ratio &  90 & -\\
Rotor, Hub radius & 63,1.5 & m \\
Blade pitch range  & 0-90 & $^o$ \\
Pitch rate  & 8   & $^o$/s \\
Optimal pitch angle ($\beta_{opt}$)   & 0 & $^o$ \\
Optimal tip-speed ratio ($\lambda_{opt}$)  & 7.55 & -\\
Maximum power coefficient ($C_{p_{max}}$)   & 0.482 & -\\
\hline
\end{tabular}
 \label{Table4}
 \end{table}
 
 The generalized equation of motion describing the nonlinear dynamics of the 5 MW NREL RWT in OpenFAST is expressed as
 \begin{equation}
   M(q,u,t)\ddot{q} + f(q,\dot{q}, u,u_d,t)=0,
   \label{eqn1}
 \end{equation}  
 where $M$ denotes the mass matrix containing inertia and mass components, $f$ the nonlinear function of the enabled DOFs $q$  and their first derivative $\dot{q}$ as well as the control input $u$, disturbance input $u_d$, and time $t$.  

 \subsection{$\mu$-Synthesis robust disturbance accommodating controller}
 The CPC-based $\mu$-synthesis RDAC controller used in this work as a lifetime controller for the proposed hybrid prognosis scheme for lifetime control of WTs is described. The design approach for this controller is similar to that of the IPC-based RDAC controller proposed in previous work \cite{Kipchirchir2024}. It is briefly repeated here for principal understanding. The controller is designed to regulate the generator speed/power and reduce tower loads in the 5 MW NREL RWT. The process involves designing a generalized state-space system made of a nominal WT model, actuator dynamics, weighting functions, and an observer-based control structure with tunable gains. Futhermore, an uncertainty description of the WT system is modeled and used for control design based $\mu$-synthesis approach for improved robustness.
 
 The nonlinear model (\ref{eqn1}) is linearized in OpenFAST to obtain a reduced-order linear time-invariant (LTI) model used for designing the proposed controller.  Linearization is carried out around a steady-state operating point in the above-rated WT operation, defined by a 18 m/s constant wind speed, 12.1 rpm rated rotor speed, and 14.6 $^o$ blade pitch angle. To capture relevant dynamics in the LTI model corresponding to the desired closed-loop performance, six DOFs are enabled for linearization. These include, first tower F-A bending mode, variable speed generator, first blade F-W bending modes, and drive-train rotational flexibility. The reduced-order model obtained after linearization is expressed as
 \begin{equation}
 \begin{aligned}
 & \dot{x}=Ax + Bu + B_dd,\\
 & y= Cx,\\
 \end{aligned}
 \label{eqn2}
 \end{equation}
 where $A\in \mathbb{R}^{11x11}$, $B\in \mathbb{R}^{11x1}$, $B_d\in \mathbb{R}^{11x1}$, and $C\in \mathbb{R}^{2x11}$ denote the system, input, disturbance, and output matrices, respectively. The measurements $y$ include generator speed $\omega_g$ ad tower-top F-A translational acceleration $\gamma$. The dynamic states $x$ include the enabled DOF displacements and velocities. The linear time invariant (LTI) model (\ref{eqn2}) is used for RDAC controller design.
 
Dynamic fluctuation in the wind field across the rotor acts as a disturbance to a WT. To counteract this effect, a disturbance accommodating controller (DAC) is designed by augmenting the LTI model (\ref{eqn2}) with an assumed wind disturbance model \cite{Wright2004}. The spatial variation wind speed is assumed to be an additive disturbance having its waveform model expressed as
\begin{equation}
\begin{aligned}
& d=\theta x_d, \\
& \dot{x}_d= Fx_d,
\end{aligned}
\label{eqn3}
\end{equation}
where $x_d$ denotes the wind disturbance state and the disturbance state-space model is $\theta$ and $F$. With a step waveform in which $\theta$ = 1 and $F$ = 0, a close estimate of sudden uniform fluctuations in horizontal wind speed is obtained \cite{Wright2008,Fingersh2004}. Combining this with high DAC gains yields a suitable solution for cancelling the disturbance \cite{Soeffker1995}. Therefore, augmenting the LTI model (\ref{eqn2}) with the wind disturbance model yields an extended linear model expressed as
  
  \begin{equation}
   \begin{split}
   \underbrace{
   \begin{bmatrix}
   \dot x\\
   \dot x_d
   \end{bmatrix}}_{\dot{x}_a}
  &=
   \underbrace{
   	\begin{bmatrix}
   	A & B_{d}\theta\\
   	0 & F
   	\end{bmatrix}}_{A_a}  
   \underbrace{
   	\begin{bmatrix}
   	x\\
   	x_d
   	\end{bmatrix}}_{x_a}   
   +
   \underbrace{
   	\begin{bmatrix}
   	B\\
   	0
   	\end{bmatrix}}_{B_a}
   u,\\
   y&=
  \underbrace{
   	\begin{bmatrix}
   	C&0
   	\end{bmatrix}}_{C_a}
   \begin{bmatrix}
   x\\
   x_d
   \end{bmatrix}.
   \end{split}
   \label{eqn4}
   \end{equation}
   
To implement full-state feedback control, an extended observer used for estimating system and disturbance states is designed assuming full observability of model (\ref{eqn4}). The observer-based system is expressed as
\begin{equation}
\begin{aligned}
& \dot{\hat{x}}_a=(A_a+B_au-LC_a)\hat x_a+Ly, \\
& \hat y=C_a \hat x_a,
\end{aligned}
\label{eqn5}
\end{equation}
where $L$ denotes the observer gain matrix, $u=-K_a\hat x_a$ the control input signal, $\hat x_a$ and $\hat y$ the estimated states and measured outputs, respectively. The controller gain matrix $K_a$ is used to cancel wind disturbances, regulate generator speed, and reduce tower F-A loads. Furthermore, to avoid steady-state errors in regulating generator speed $\omega_g$, model (\ref{eqn5}) is extended with an integral state $x_i=\int \omega_gdt$, yielding a partial integral action $\dot{x}_i=C_iy$ with $C_i $ denoting the location of $\omega_g$. Therefore, observer-based DAC model with integral action becomes
 \begin{equation}
  \begin{split}
  \underbrace{
  \begin{bmatrix}
  \dot{\hat{x}}_a\\
   \dot x_i
  \end{bmatrix}}_{\dot x_b} 
  &=
  \underbrace{
  	\begin{bmatrix}
  	A_a-B_aK_a-LC_a & B_aK_i\\
     0 & 0\\
  	\end{bmatrix}}_{A_b} 
  \underbrace{
  	\begin{bmatrix}
  	\hat x_a\\
  	\hat x_i
  	\end{bmatrix}}_{x_b} 
  +
  \underbrace{
  	\begin{bmatrix}
  	L\\
  	C_i
  	\end{bmatrix}}_{B_b}
  y,\\
  u&=
  \underbrace{
  	-\begin{bmatrix}
  	K_a & K_i
  	\end{bmatrix}}_{C_b}
  \begin{bmatrix}
  \hat{x}_a\\
  \hat{x}_i
  \end{bmatrix}  
  ,
  \label{eqn6}
  \end{split}
  \end{equation} 
 where $K_i$ denotes the integral gain. Classical approaches such as linear quadratic regulator (LQR) and pole placement can be used for designing the gains $K_a$ and $K_i$ separately. Therefore, full-system stability, robustness, and optimality is not considered. In addition, because uncertainties are not considered in the design process, modeling errors are introduced. This can potentially affect closed-loop robustness. In this work, $\mu$-synthesis approach is used for designing these gains in a single step. This guarantees the desired robust performance in tower load mitigation and generator speed regulation.
 
 To optimize the parameters of the structured DAC controller (\ref{eqn6}), the H$_\infty$ control approach (\ref{eqn7}) is used as a cost function for. The find an optimal RDAC controller $RDAC^*(L,K)$ defining the optimal gains $K=[K_a \text{ } K_i]$ and $L=[L \text{ } C_i]^T$ is formulated as an optimization problem expressed as
 \begin{equation}
 RDAC^*=\underset{RDAC \in \mathcal{RDAC}}{argmin}\parallel G_{zd}(P,RDAC) \parallel _\infty, 
 \label{eqn7}
 \end{equation}
 where $\mathcal{RDAC}$ denotes a set of controllers $RDAC$ that stabilize the generalized plant $P$. Additionally, to guarantee asymptotic stability of the closed-loop system, the optimization problem (\ref{eqn7}) is subjected to Lyapunov stability constraint given as $\parallel C_b(sI-\mathcal{A}(RDAC))^{-1}B_b \parallel _\infty<+\infty,$ where $\mathcal{A}(RDAC)$ denotes the system matrix $A_b$.

\begin{figure*}[thpb]
 \centering
 \includegraphics[width=0.8\linewidth]{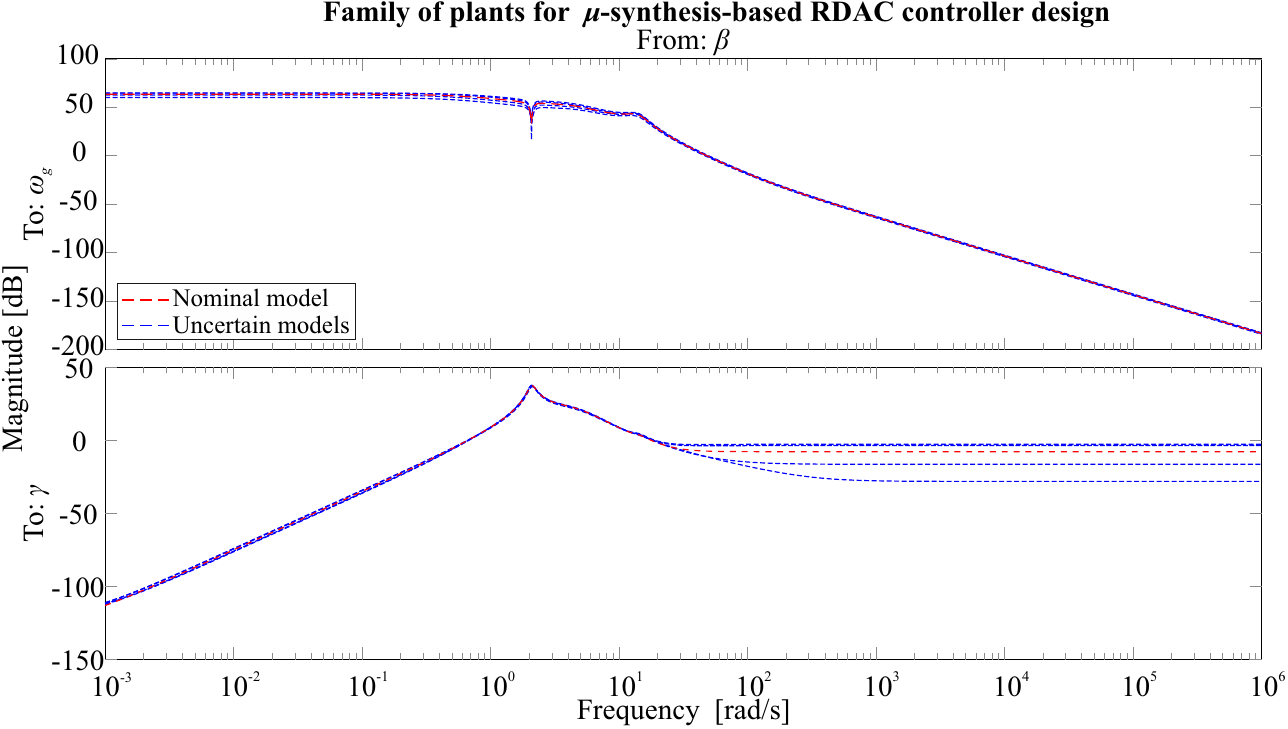}
 \caption{Comparison of open-loop frequency responses from collective pitch angle to measured outputs} 
 \label{fig3}
 \end{figure*}

 The nominal model (\ref{eqn2}) defines the WT dynamics at a specific operating point defined by a hub-height wind speed of 18 m/s. In changing operating conditions, model uncertainties are expected to manifest as WT dynamics deviate from the model dynamics. To illustrate this, the open-loop frequency response of a sample of linear models obtained at different operating points (wind speeds of 14 m/s, 16 m/s, 20 m/s, and 22 m/s) is obtained as shown in Fig. \ref{fig3}. It can be seen that there is a noticeable variation in the responses between the nominal and uncertain models, particularly for the tower F-A acceleration at high frequencies above the rated rotor speed of 1.267 rad/s. Therefore, a description of these uncertainties should be included in controller design. Due to the modest variability in the frequency response of the uncertain models, the wind disturbance is assumed to produce an additive uncertainty.
 
 The family of linear models (Fig. \ref{fig3}) is used to model an unstructured additive uncertainty description. Therefore, the frequency-dependent unstructured uncertainties resulting from unmodeled dynamics in the linearization process are included in the nominal plant model (\ref{eqn2}). In this way, an uncertain plant model  $\tilde{G}$ of the form $\tilde{G}=G+W\Delta_a$ is obtained and used for controller design. Here, $\Delta_a$ denotes the uncertain dynamics with unit peak gain and $W$ denotes a 2$\times$2 diagonal shaping filter. The orders of individual diagonal elements are designed to adjust the degree of uncertainty at each frequency. This ensures that the gaps between the nominal and uncertain models are closely tracked, improving uncertainty estimation. In this thesis, the orders of each diagonal entry of $W$ are designed to shape the uncertainty in the respective outputs $y$ of the family of uncertain plants. The obtained uncertainty  $\Delta=W\Delta_a$, which has a $2\times2$ block diagonal structure is used for designing the proposed RDAC controller. Structured or parametric uncertainties, usually present at low frequencies as a result of plant perturbations \cite{Geyler2008} are not considered in this thesis. This is because in above-rated WT operation, variations in parameters of interest are small because the generator speed and power are regulated to their rated values.
 
\begin{figure}[thpb]
  \centering
  \includegraphics[width=1\linewidth]{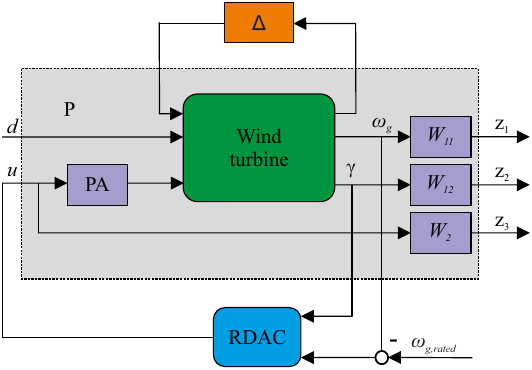}
  \caption{The $\mu$-synthesis RDAC controller applied to the 5 MW NREL RWT}
  \label{fig4}
  \end{figure}

 The proposed RDAC controller is applied to the 5 MW NREL RWT as shown in Fig. \ref{fig4}. The additive uncertainty $\Delta$ describes the variations between the nominal and uncertain family of plants in Fig. \ref{fig3}. Wind disturbance $d$ excites the WT dynamics in above-rated operation. The generalized plant P, consisting of the nominal WT plant, pitch actuator (PA), and weighting functions $W_11$, $W_11$, and $,W_11$, is interconnected with the observer-based RDAC system (\ref{eqn6}) using lower LFT. On the other hand, upper LFT is used to interconnect P and $\Delta$. The controlled outputs z = [z$_1$ z$_2$ z$_3$] include the weighted measurements $w_g$ and $\gamma$ and control input $u$, respectively.  
 
 A second order transfer function used to model PA describes the slow blade pitch actuator dynamics. It is expressed as
 
\begin{equation}
 \beta=\frac{\omega^2_{PA}}{s^2+2\zeta\omega_{PA}s+\omega^2_{PA}}\beta_{com},
 \label{eqn9}
\end{equation}
where $\beta_{com}$ and $\beta$ denote the commanded and actual pitch angles, respectively. According to the recommendation by NREL \cite{Rinker2018}, the natural frequency $\omega_{PA}$ is chosen to be four times the turbine's rated rotor speed $\omega_r$=1.267 rad/s, and the damping ratio $\zeta$ is 80 \% critical.

Using the $DK$-iteration process, the optimal RDAC is designed using $\mu$-synthesis approach by minimizing structured singular value (SSV) $\mu$. The RDAC controller relies on generator speed $\omega_g$ and tower F-A translational acceleration $\gamma$ to generate a CPC control input $u$ for regulating generator speed to its rated value $\omega_{g,rated}$ and for damping the first-mode tower F-A vibration. 
 
\begin{figure*}[thpb]
\centering
\includegraphics[width=0.8\linewidth]{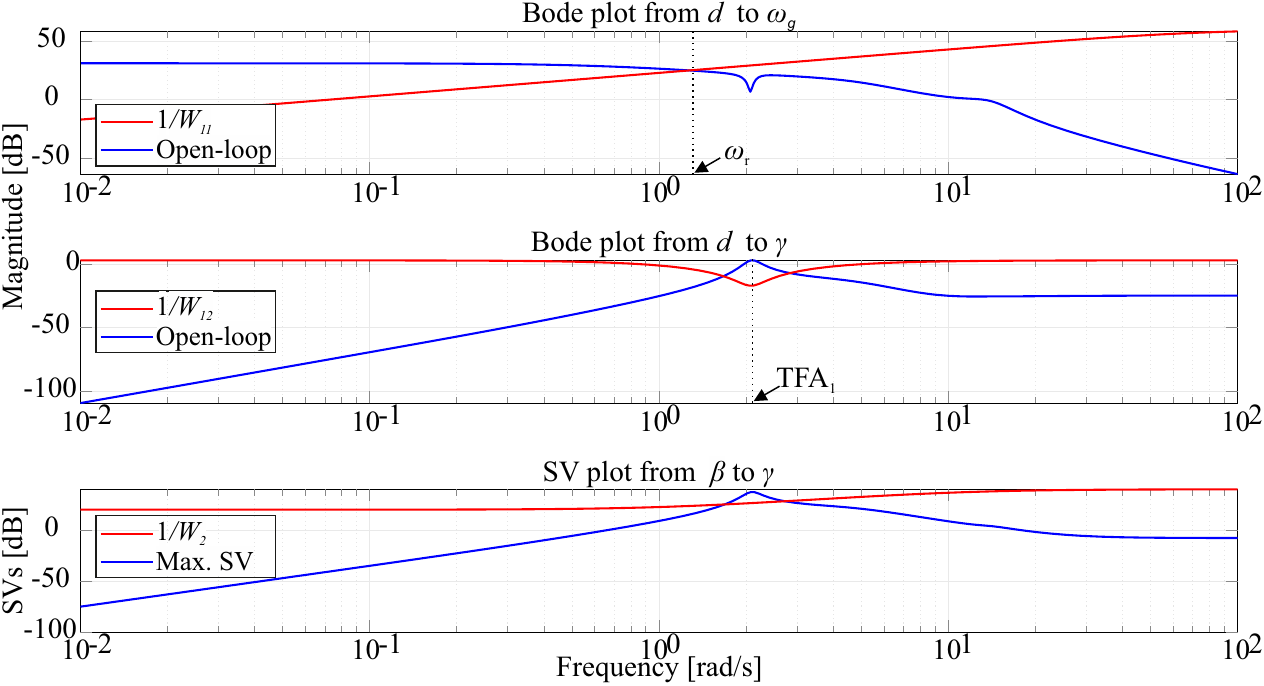}
\caption{Comparison of open-loop frequency responses and weighting functions used in RDAC control design} 
\label{fig5}
\end{figure*}

Robust performance is dictated by appropriate selection of weighting functions. Closed-loop characteristics are shaped using the desired weighting functions that are rational, stable, and minimum phase \cite{Skogestad2005}. Therefore, the frequency-dependent weighting functions $W_{11}$, $W_{12}$, and $W_2$ are designed to shape system measurement signals and control input to achieve the desired closed-loop frequency response. In Figure \ref{fig5}, Bode diagrams of the open-loop transfer functions from wind disturbance $d$ to measurement outputs compared with associated inverted weighting functions $1/W_{11}$ and $1/W_{12}$ are shown. A singular value (SV) plot from control input $\beta$ to the tower F-A acceleration $\gamma$ compared with $1/W_2$ is also shown. As illustrated, $1/W_{11}$ and $1/W_{12}$ shape the respective open-loop responses to achieve the desired closed-loop frequency responses. To effect the generator speed response while ensuring robustness to wind disturbances, $W_{11}$ is designed as an inverted LPF. To damp the first tower F-A vibration mode, $W_{12}$ is designed as an inverted notch filter centred at this frequency TFA$_1$=2.08 rad/s. To reduce controller activity in high frequencies and thus increase robustness, $W_2$ is designed as an inverted LPF, with control bandwidth being limited to effect the desired frequency TFA$_1$.

The upper LFT $F_u(M,\Delta)$ in Fig. \ref{fig4} consists of the transfer function $M$ from the output to the input of the perturbation $\Delta$, both of which are stable. The lower LFT $N=F_l(P,RDAC)$ interconnects the plant $P$ with the observer-based $RDAC$ controller. For robust stability (RS), an RDAC controller can be obtained if the system remains stable for all uncertain plants shown in Fig. \ref{fig3}. Robust performance (RP) is guaranteed if RS condition is met and in addition, the performance objective can be achieved for all possible plants in the uncertainty set, including the worst-case plant. These criteria are expressed as
\begin{equation}
RS \Leftrightarrow F_u(M,\Delta) \text{ }\textnormal{is stable for} \text{ } \forall\Delta,\parallel\Delta\parallel_\infty \le 1; \textnormal{and NS},
\label{eqn9}
\end{equation}
\begin{equation}
RP \Leftrightarrow \parallel F_l(N,\Delta) \parallel_\infty < 1 \text{ }\textnormal{for} \text{ } \forall\Delta,\parallel\Delta\parallel_\infty \le 1; \textnormal{and NS},
\label{eqn10}
\end{equation}
where NS denotes nominal stability. 

From small gain theorem, RS is expressed as
\begin{equation}
RS \Leftarrow \bar{\sigma}(M) < 1 \text{ }\forall\omega,
\end{equation}
which is a tight condition for any case of complex $\Delta$ satisfying $\bar{\sigma} \le 1$. A more general tight condition is given as
\begin{equation}
RS \Leftrightarrow\mu(M) < 1 \text{ }\forall\omega,
\end{equation}
where the real non-negative $\mu(M)$ is the SSV which is expressed as
\begin{equation}
\mu(M)=\frac{1}{\min\{k_m|\det(I-k_mM\Delta)=0 },
\end{equation}
for structured $\Delta$, $\bar{\sigma} \le 1$. If no structured uncertainty $\Delta$ exists then $\mu(M)=0$. The factor $k_m$ is used to scale the uncertainty $\Delta$ to make the matrix $I-k_mM\Delta$ singular, hence the SSV is expressed as $\mu(M)=1/k_m$ \cite{Skogestad2005}.

Therefore, $\mu$-synthesis is a important tool for RP analysis. To synthesize a controller to minimize a given $\mu$ condition, a scaled version of $\mu$ is needed \cite{Mirzaei2011}. To synthesize an optimal RDAC controller that minimizes $\mu$ by guaranteeing RP and RS, the $DK$-iteration process is used in this thesis. The process solves a sequence of scaled H$_\infty$ problems by using frequency-dependent scaling matrices, $D$ and $G$, which take advantage of the uncertainty structure. First, nonsmooth H$_\infty$ synthesis \cite{Apkarian2017} is used to obtain an RDAC controller that minimizes the closed-loop gain of the nominal plant P. The process is summarized as
\begin{equation}
\min_{RDAC}(\min_{D\in\mathcal{D}}\parallel DN(RDAC)D^{-1}\parallel_\infty).
\end{equation}

In the $D$-step, the robust H$_\infty$ performance of the closed-loop system using the current RDAC controller is estimated. The upper bound $\bar{\mu}$ of the robust H$_\infty$ performance of RDAC is then computed using a suitable $D(j\omega)$ scaling which commutes with $\Delta$. Rational functions $D(s)$ of a specified order are used to fit the $D(j\omega)$ scaling, yielding the scaled H$_\infty$ norm $\mu_F$. In the $K$ step, a controller $RDAC^*$ that minimizes $\mu_F$ to improve the robust performance obtained in the $D$-step is synthesized. The iterative process is repeated until no further improvement in $\mu$ is achieved by the optimal controller $RDAC^*$. In this work, only $D$ scaling is used because wind disturbance is assumed to produce a complex additive uncertainty. 
 
\subsection{Online Damage Evaluation}
Linear damage accumulation theory based on Miner's rule is widely used due to its simplicity. Wind speed variability produces a complex spectrum of varying loads in WT components. Therefore, to use Miner's rule, the RFC algorithm is used to transform this spectrum into simple load cycles. Stress range histograms can be extracted and used for evaluating damage accumulation $D_k$ as follows
 
 \begin{equation}
   D_k=\sum_{i=1}^{k}d_i=\sum_{i=1}^{k}\frac{n_i}{N_i}=\sum_{i=1}^{k}\frac{n_is_i^m}{K},
     \label{eqn11}
 \end{equation}
 where $k$ denotes the total number of related stress range histograms, $d_i$ the incremental damage at the $i^{th}$ stress range histogram, $n_i$ the number of applied load cycles in each histogram bin, $N_i$ the number of cycles to failure at the $i^{th}$ stress range histogram, and $s_i$ the applied load amplitude in each histogram bin.
 
 Standard RFC algorithms generate equivalent load cycles from complex load spectra by pairing local minimum and maximum points using the three-point counting rule. However, they cannot be used for online monitoring or lifetime control as this requires the entire load history beforehand for the load cycles to be generated \cite{Musallam2012}. To solve this, a real-time implementation of the RFC algorithm is proposed in \cite{Musallam2012}. By employing a three-point recursive counting rule, the extremal points of timeseries loading data are processed online and stored in two flexible stacks to identify the full and half cycles. Using Miner’s rule for each identified cycle, the life consumption of a component is calculated and incremented online, hence allowing for real-time determination of the consumed life of a component. 
 
 In this work, the online damage evaluation model \cite{Musallam2012} is used to evaluate the tower damage accumulation at every timestep $D_k$.  The estimated lifetime of the tower $L_e$=600 seconds, used as the SOH indicator, is calculated
 \begin{equation}
  L_e= \frac{T_k}{D_k}D_d,
  \label{eqn12}
 \end{equation}
where $T_k$ denotes the time at the current time step while $D_d$ denotes the accumulated tower damage at the defined lifetime.
 
 \subsection{Online Load Prediction}
\begin{figure*}[thpb]
  \centering
  \includegraphics[width=1\linewidth]{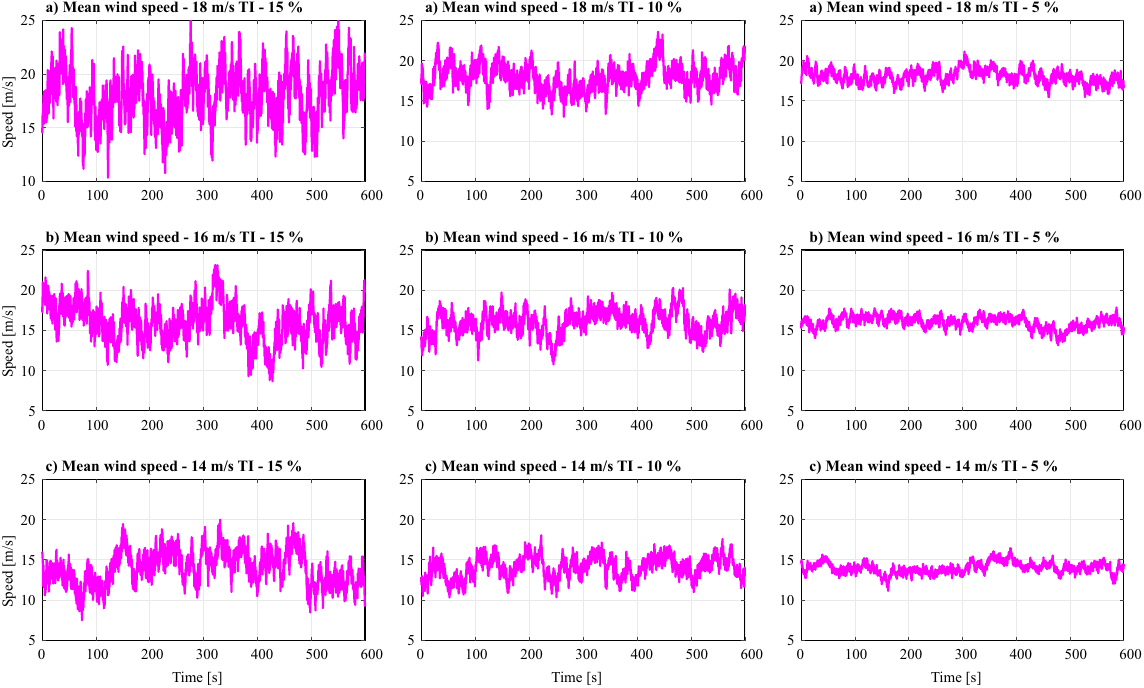}
  \caption{ Wind profiles used for SVM regression training and testing}
  \label{fig6}
  \end{figure*}

 Due to the massive size of the towers in commercial WTs, strain gauge measurements are unreliable and are usually not used for load determination. However in OpenFAST software \cite{OpenFAST2021}, it is provided for design purposes. Therefore, to implement prognosis of tower fatigue life in an actual WT, its load should be estimated using available measurements. In this thesis, an SVM regression model is developed for predicting tower F-A bending moment. The WT measurements used as predictors include, the horizontal hub-height wind speed $d$, tower F-A displacement $\sigma$, rotor power $p$ , and tower F-A acceleration $\gamma$. For training and testing, timeseries data from closed-loop simulations in various wind conditions shown in Fig. \ref{fig6} is used. Six wind profiles having TIs of 5 \% and 15 \% are used for training, and three with 10 \% TI are used for testing. In the training phase, the best set of hyperparametes is obtained by applying Bayesian optimization. To evaluate the performance of the model, root mean square error (RMSE) is used, with the model using a linear kernel function returning the lowest RMSE. For training and testing, the RMSE of the predicted response is 609~kNm and 597.3~kNm, respectively, corresponding to an accuracy of 98.4 \% and 98.5 \%, respectively. The high prediction accuracy is validated using a new set of wind profiles for simulating the proposed prognosis scheme. It is important to note that other ML regression models such as Gaussian process regression (GPR) and a neural network (NN) were used for training. Although GPR model returned the same prediction accuracy as SVM regression, the training time was quite long. On the other hand, although the NN regression model had slightly better prediction accuracy, it had slow convergence, which is not suitable for online implementation.
 
\subsection{A Hybrid Prognosis-based Robust Lifetime Control: An Illustrative Example Using The 5 MW NREL WT Model}
\begin{figure}[thpb]
  \centering
  \includegraphics[width=1\linewidth]{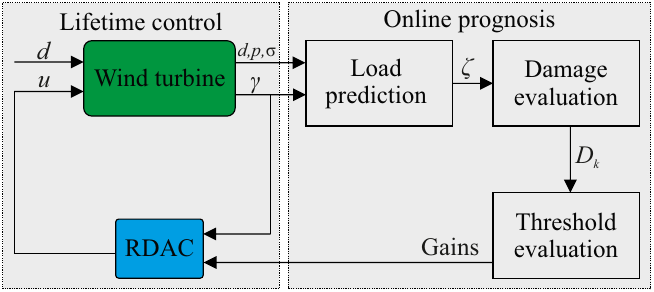}
  \caption{The hybrid lifetime prognosis scheme applied to the 5 MW NREL WT model}
  \label{fig7}
  \end{figure}

The proposed hybrid prognosis approach applied for controlling the lifetime of the 5 MW NREL RWT using RDAC controller is shown in Fig. \ref{fig7}. The online prognosis scheme is realized using an SVM regression model for tower F-A load prediction $\zeta$ and a damage evaluation model for real-time calculation of $D_k$. The threshold evaluation model uses $D_k$ to calculate $L_e$ using (\ref{eqn12}). Based on the threshold values set for $L_e$, RDAC controller gains are adapted continuously to vary the degree of trade-off between tower load mitigation and generator speed regulation. This ensures that the predefined damage limit is not exceeded at the desired lifetime and that generator speed regulation is not compromised.

\section{Results and discussion}
\label{sec:results}
To validate the proposed hybrid prognosis scheme for lifetime control, closed-loop dynamic simulations are performed in OpenFAST using the 5 MW RWT. The goal of the prognosis scheme is to control the fatigue life consumption of the tower while maintaining optimal generator speed regulation. The performance of the hybrid prognosis scheme, henceforth denoted as Life2, is evaluated against a none-lifetime RDAC controller, henceforth denoted as Baseline, tuned for optimal trade-off between tower load mitigation and generator speed regulation. It is also compared with the model-based prognosis scheme, henceforth denoted as Life1, which uses RDAC as the lifetime controller and relies on actual tower load measurements for fatigue damage evaluation using online damage accumulation model \cite{Musallam2012}.

\subsection{Lifetime Control}
\begin{figure*}[thpb]
  \centering
  \includegraphics[width=1\linewidth]{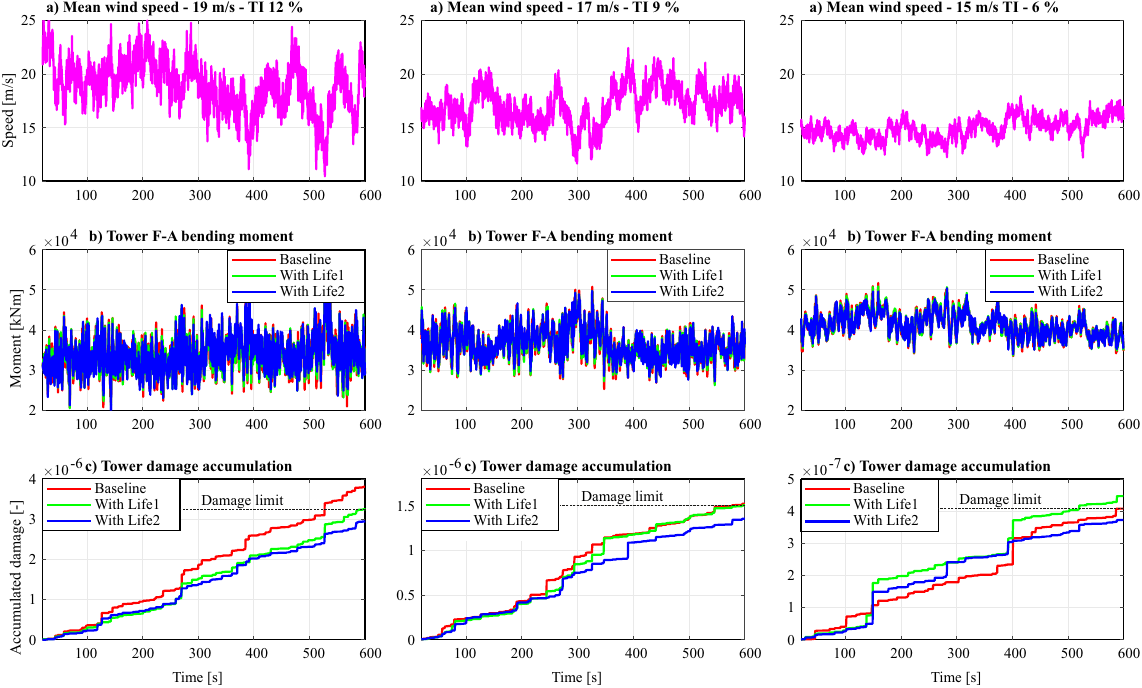}
  \caption{Tower F-A load response and damage accumulation}
  \label{fig8}
  \end{figure*}

A new set of wind profiles shown in Fig. \ref{fig8}a are used in the simulations. In Figure \ref{fig8}b, the tower F-A load response is shown. he tower F-A load response is shown. The baseline controller exhibits the highest vibrations compared with the lifetime prognosis-based control schemes Life1. The proposed hybrid prognosis scheme Life2 shows similar responses to Life1, hence validating the high prediction accuracy of the SVM regression model. In Table \ref{Table5}, the $\delta$ in tower F-A load is shown. Both lifetime controllers achieve 3.4 \% improvement in load reduction. In Figure \ref{fig8}c, the tower damage accumulation is shown. In both 19 m/s and 17 m/s wind fields, the lifetime controllers achieve the desired damage limit at the predefined lifetime of 600 s. However, for the 15 m/s wind profile, Life1 does not achieve the damage limit due to the suboptimal trade-off in control objectives at this operating point, which can be improved by retuning the RDAC controller. On the other hand, the proposed hybrid scheme meets the desired damage limit.

  \begin{table*}[h!]
  \centering
  \caption{Load mitigation, pitch activity, and generator speed and power regulation performance analysis (Key: \textcolor{green}{best}, \textcolor{red}{worst})}
  \begin{tabular}{llllllll}
  
  \textbf{Parameter} & \textbf{Units}	& \textbf{Control scheme}& \textbf{19 m/s} & \textbf{17 m/s} & \textbf{15 m/s} & \textbf{Avg.} & \textbf{\%}\\
  \hline
    Tower F-A ($\delta$)  & kNm & Baseline & \textcolor{red}{4438.3}  & \textcolor{red}{3855.1}  & \textcolor{red}{3028.4}   & \textcolor{red}{3773.9} & \textcolor{red}{-} \\
                       & & Life1 & \textcolor{black}{4265.6}  & \textcolor{black}{3709.0}  & \textcolor{black}{2963.0} & \textcolor{black}{3645.9} & \textcolor{black}{-3.39}  \\
                       & & Life2 & 4265.6 & \textcolor{green}{3708.7}  & \textcolor{black}{2963.0} & \textcolor{green}{3645.8} & \textcolor{green}{-3.40} \\
    \hline
   Pitch rate (RMS)   & $^o$/s        & Baseline & \textcolor{green}{3.946}  & \textcolor{green}{2.152}  & \textcolor{green}{0.976}  & \textcolor{green}{2.358} & \textcolor{green}- \\
                      & & Life1 & \textcolor{black}{5.500}  & \textcolor{black}{2.955}  & \textcolor{black}{1.208} & \textcolor{black}{3.221} & \textcolor{black}{36.6}  \\
                      & & Life2 & \textcolor{black}{5.500}  & \textcolor{red}{2.959}  & \textcolor{black}{1.208} & \textcolor{red}{3.223} & \textcolor{red}{36.7} \\
  \hline
    Generator speed (RMSE) & rpm      & Baseline & \textcolor{black}{12.14}  & \textcolor{red}{9.46}  & \textcolor{black}{6.46}   & \textcolor{black}{9.35} & \textcolor{red}- \\
                     & & Life1 & \textcolor{green}{12.05}  & \textcolor{green}{9.33} & \textcolor{green}{6.32} & \textcolor{green}{9.23} & \textcolor{green}{-1.26}  \\
                    & & Life2 & \textcolor{black}{12.14}  & \textcolor{black}{9.45}  & \textcolor{black}{6.46} & \textcolor{black}{9.35} & \textcolor{black}{-0.01} \\                                  
  \hline
  Generator power (RMSE) & kW      & Baseline & \textcolor{black}{166.46}  & \textcolor{black}{106.45}  & \textcolor{black}{55.95}   & \textcolor{black}{136.46} & \textcolor{black}- \\
                     & & Life1 & \textcolor{red}{177.87}  & \textcolor{red}{108.10} & \textcolor{red}{56.38} & \textcolor{red}{142.98} & \textcolor{red}{4.78}  \\
                    & & Life2 & \textcolor{black}{166.46}  & \textcolor{green}{106.43}  & \textcolor{black}{55.95} & \textcolor{green}{136.45} & \textcolor{green}{-0.01} \\    
  \hline
  \end{tabular}
  \label{Table5}
  \end{table*}
\subsection{Generator Speed Regulation}
\begin{figure*}[thpb]
  \centering
  \includegraphics[width=1\textwidth]{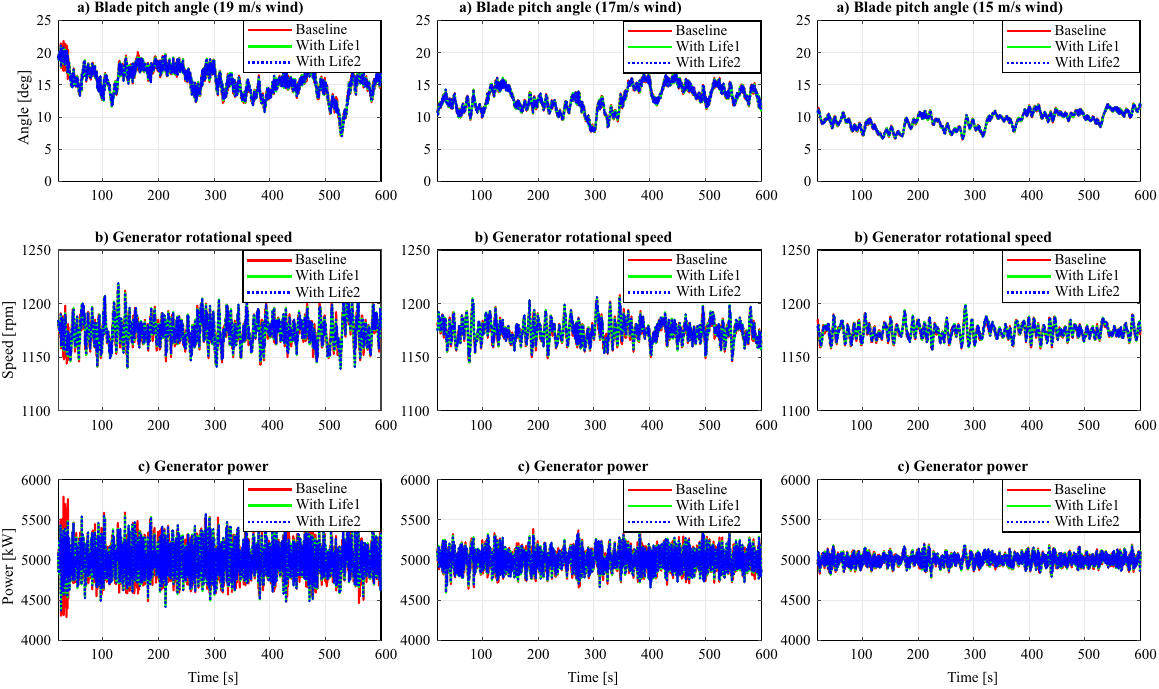}
  \caption{Generator speed and power regulation}
  \label{fig9}
  \end{figure*}

The impact of applying the proposed hybrid prognosis scheme on generator speed and power regulation is evaluated as shown in Fig. \ref{fig9}. To reduce tower fatigue damage, the proposed scheme Life2 increases pitch activitity as illustrated in Fig. \ref{fig9}a. However, as shown in Table \ref{Table5}, RMSE of pitch rate (PR) does not exceed the maximum PR constraint of 8 $^o$/s for the 5 MW NREL RWT. Compared with the baseline controller, Life2 ensures optimal generator speed and power regulation as shown in Fig. \ref{fig9}b and the RMSE values in Table \ref{Table5}. However, Life1 shows better speed regulation. 

\section{Summary and conclusion}
\label{sec:conclusion}
In this contribution, a hybrid prognosis scheme for robust lifetime control of WTs is presented. To realize online prognosis, an SVM regression model is developed for tower load prediction, and a real-time RFC damage evaluation model is used for SOH estimation. The estimated lifetime of the tower is evaluated at each time-step. Based on a set of lifetime thresholds, the gains of the $\mu$-synthesis RDAC controller are adapted, ensuring that the damage limit of the component is not exceeded at the desired lifetime. The 5 MW NREL RWT is used to evaluate the effictiveness of the proposed approach. Its performance is compared with a model-based lifetime control scheme, which uses ideal WT load measurements. Dynamic simulation results show that the proposed approach controls the life consumption of the WT tower to achieve a predefined damage limit and lifetime without compromising generator speed regulation. The perfomance of the proposed approach is comparable to the model-based lifetime control method and is therefore suitable for practical implementation. The limitation of the proposed approach is that a linear degradation model is used for a fatigue load estimation. However, WT components experience complex nonstationary degradation patterns due to several failure modes. State-of-the-art (SOTA) model-based prognosis approaches such as Bayesian method (BM) or particle filter (PF) can be considered for improve fatigue life estimation. Furthermore, due to abundant data collected from modern high frequency SCADA systems, deep learning ML regression models can be considered for improved load prediction accuracy. Therefore by combining physics-based models using SOTA methods such as BM or PF, and deep learning algorithms, hybrid prognostics-based robust lifetime control approaches can be developed for improved RUL prediction. 

\section*{Acknowledgment}
The authors acknowledge support by the Open Access Publication Fund of the University of Duisburg-Essen, Germany. They also acknowledge the support granted to the first author by German Academic Exchange Service (DAAD) in cooperation with the Ministry of Education of Kenya, for his Ph.D. study at the Chair of Dynamics and Control, UDE, Germany.

\bibliographystyle{IEEEtran}
\bibliography{paper_vf}

\begin{IEEEbiography}[{\includegraphics[width=1in,height=1.25in,clip,keepaspectratio]{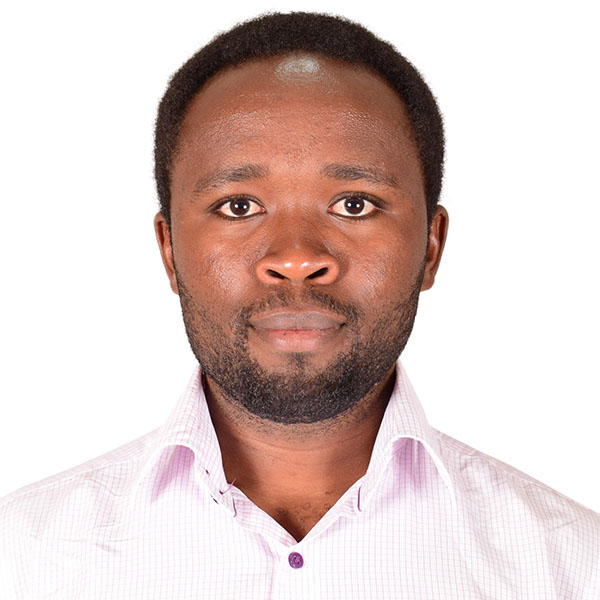}}]{Edwin Kipchirchir} received his B.S. and M.S. degrees in mechatronic engineering from Jomo Kenyatta University of Agriculture and Technology, Nairobi, Kenya, in 2013 and 2019, respectively. He is currently pursuing a Ph.D. degree at the Chair of Dynamics and Control, University of Duisburg-Essen, Germany. His current research interests include robust control methods applied to load mitigation and lifetime control of WT structures. He currently focuses on developing advanced control schemes for structural load mitigation, speed regulation, and prognosis in utility-scale wind turbines.
\end{IEEEbiography}

\begin{IEEEbiography}[{\includegraphics[width=1in,height=1.25in,clip,keepaspectratio]{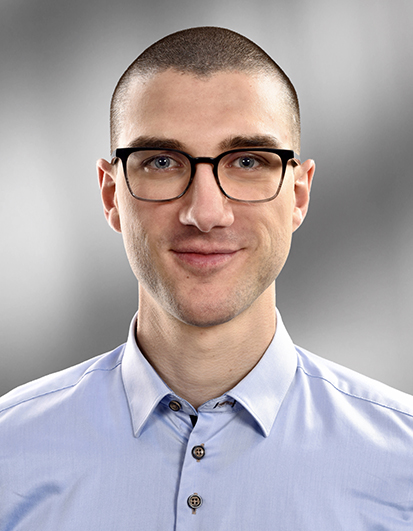}}]{Jonathan Liebeton} received his B.S. and M.S. degrees in mechanical engineering from University of Duisburg-Essen, Germany, in 2019 and 2021, respectively. He is currently pursuing a Ph.D. degree at the Chair of Dynamics and Control, University of Duisburg-Essen, Germany. His current research interests include machine learning-based structural health monitoring. He currently focuses on diagnosis of carbon fibre reinforced plastics using Acoustic Emission.
\end{IEEEbiography}

\begin{IEEEbiography}[{\includegraphics[width=1in,height=1.25in,clip,keepaspectratio]{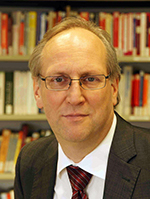}}]{Dirk S{\"o}ffker} (Member, IEEE) received his Ph.D. degree (Dr.-Ing.) in mechanical engineering and the Habilitation degree in automatic control/safety engineering from the University of Wuppertal, Germany, in 1995 and 2001, respectively. Since 2001 he has been the head of the Chair of Dynamics and Control, University of Duisburg-Essen, Germany. Since 2023 he is heading the Mechanical and Process Engineering Department and serves as Vice Dean of the Engineering Faculty responsible for the Mechanical Engineering department. His current research interests include diagnostics and prognostics, modern methods of control theory, safe human interaction with technical systems, safety and reliability control engineering of technical systems, and cognitive technical systems. 
\end{IEEEbiography}

\EOD

\end{document}